\documentclass[superscriptaddress,pra,onecolumn,aps,showpacs]{revtex4}

\usepackage{amsfonts,epsfig}
\usepackage{latexsym}
\usepackage{epsfig}
\usepackage{amsmath}
\usepackage{amssymb}
\usepackage{amsfonts}
\usepackage{amsthm}
\usepackage{mathrsfs}
\usepackage{algorithm}
\usepackage{algorithmic}
\DeclareMathAlphabet{\mathrsfs}{U}{rsfs}{m}{n}
\DeclareMathAlphabet{\mathpzc}{OT1}{pzc}{m}{it}
\DeclareMathAlphabet{\matheus}{U}{eus}{m}{n}
\DeclareMathAlphabet{\mathbbold}{U}{bbold}{m}{n}

\newcommand{\ba}{\begin{eqnarray}}
\newcommand{\ea}{\end{eqnarray}}

\newcommand{\ket}[1]{\left | #1 \right \rangle}

\newcommand{\HH}{\mathcal{H}}

\newcommand{\Tr}{\operatorname{Tr}}

\newcommand{\comment}[1]{}

\begin{document}

\title{Finite-key security against coherent attacks in quantum key distribution}
\author{Lana Sheridan}
\affiliation{Centre for Quantum Technologies, National University of Singapore, Singapore}
\author{Thinh Phuc Le}
\affiliation{Centre for Quantum Technologies, National University of Singapore, Singapore}
\author{Valerio Scarani}
\affiliation{Centre for Quantum Technologies, National University of Singapore, Singapore}
\affiliation{Department of Physics, National University of Singapore, Singapore}

\date{\today}

%______________________________________________________________________ ABSTRACT

\begin{abstract}
The work by Christandl, K\"onig and Renner [Phys. Rev. Lett. \textbf{102}, 020504 (2009)] provides in particular the possibility of studying unconditional security in the finite-key regime for all discrete-variable protocols. We spell out this bound from their general formalism. Then we apply it to the study of a recently proposed protocol [Laing \textit{et al.}, Phys. Rev. A \textbf{82}, 012304 (2010)]. This protocol is meaningful when the alignment of Alice's and Bob's reference frames is not monitored and may vary with time. In this scenario, the notion of asymptotic key rate has hardly any operational meaning, because if one waits too long time, the average correlations are smeared out and no security can be inferred. Therefore, finite-key analysis is necessary to find the maximal achievable secret key rate and the corresponding optimal number of signals.
\end{abstract}


\maketitle

%______________________________________________________________________ Article

\section{Introduction}
\label{intro}

Quantum key distribution (QKD) provides a way of distributing secret keys for use in secure communication \cite{review1,review2}. Started by Bennett and Brassard in 1984 (BB84, \cite{bb84}) and by Ekert in 1991 \cite{e91}, QKD has posed several challenges, both theoretical and experimental, which have been met to a large extent. One of those challenges has been the derivation of security bounds that take into account the finite number $N$ of exchanged quantum signals, i.e. the finite size of the keys one has to work with. The tools for such a study were remarkably anticipated by Mayers in his very first unconditional security proof \cite{may96}, but for several reasons the full solution was delayed by more than 10 years. Hayashi's formalism \cite{hay2} was tailored for the BB84 protocol. The approach by Renner and one of us \cite{SR08,CS09,SS10} is in principle more flexible but is limited to collective attacks in general: unconditional security could be claimed only for BB84 and those other few protocols, in which the bound for collective attacks is known to coincide with the one for the most general attacks \cite{KGR05}. Recently, Christandl, K\"onig and Renner developed some very general mathematical tools \cite{CKR09}, one of whose applications is the derivation of finite-key bounds for any discrete-variable protocol (for the status of the question in continuous-variable protocols  see \cite{LGG10}).

In this paper, we spell out explicitly the method to compute the finite-key QKD bound described in \cite{CKR09}. This new tool can be used to compute unconditional security bounds in the finite-key regime for protocols like Bennett 1992 (B92 \cite{b92}), Scarani-Ac\'{\i}n-Ribordy-Gisin 2004 (SARG04 \cite{sarg04,BGKS05}) or protocols based on the violation of Bell's inequalities \cite{AMP06,SR08b}. As an application, we have rather chosen the \emph{reference frame independent} protocol proposed by Laing \emph{et al.}~\cite{LSROB10}. This protocol is useful in situations, in which the alignment of reference frames between Alice and Bob is not monitored and may vary in time. In this study we consider the finite key analysis of this protocol, in light of the fact that the reference frames relations in these scenarios will not only be unknown, but may also be fluctuating over the course of the protocol. Under these assumptions, one must find that optimal secret key rates are reached for a finite number of signals: if Alice and Bob wait too long time, their correlations will be smeared due to the misalignment of the frames.

The paper is arranged as follows. In Section~\ref{method} we present the new method for finite key analysis against coherent attacks. In Section~\ref{application}, we use this method to analyze the reference frame independent protocol for two cases of drifting phase references: firstly, one frame rotating at constant speed relative to the other; secondly, the angle between the frames fluctuating according to a random walk. Lastly, in Section~\ref{conclusion} the implications of the results are considered.

\section{Finite Key Analysis Method}
\label{method}

We start by summarizing the notations and the bound for collective attacks, as discussed in detail in previous works \cite{SR08,SR08b,CS09,SS10}. Then we present the new bound extracted from \cite{CKR09}. 

\subsection{Notations and bound for collective attacks}

Let $N$ be the number of signals sent by Alice that are received by Bob. In addition to the error rate in the raw key, denoted $Q$, the protocol uses $n_{\text{PE}}$ parameters $\mathbf{V}=\{v_1,...,v_{n_{\text{PE}}}\}$ to bound Eve's information. For simplicity, we consider asymmetric protocols~\cite{LCA}, in which $n$ signals are used to create the raw key, while other signals are used to estimate the other parameters (the secret key rate for the symmetric protocol is larger by $n_{\text{PE}}+1$ at most and becomes the same in the asymptotic limit). The number of signals devoted to estimating $v_j$ is written $m_j$.

Let now $\varepsilon_{PA}$ be the probability that privacy amplification fails, and $\varepsilon_{\text{PE}}$ the probability that the real value of a parameter lies outside of the chosen fluctuation range. There is a third error probability, denoted $\bar{\varepsilon}$, which measures the accuracy of estimation of the smooth min-entropy. Finally, there is a probability $\varepsilon_{EC}$ that error correction fails, which is determined by the choice of the error correction code. Because of the \textit{composability} of the bound, in the worst case, the probability $\varepsilon_{\text{col}}$ that the quantum key distribution protocol fails does not exceed the sum of the probabilities of failure in different phases of the protocol:
\begin{equation}
\label{eq:epscollective}
\varepsilon_{\text{col}} = \varepsilon_{\text{PA}} + \bar{\varepsilon} + n_{\text{PE}}\varepsilon_{\text{PE}} + \varepsilon_{\text{EC}}.
\end{equation} The user can choose $\varepsilon_{\text{col}}$ and $\varepsilon_{\text{EC}}$; the other parameters can be optimized under the constraint (\ref{eq:epscollective}).

If the key alphabet is made by $d$-valued symbols, the secret key fraction against collective attacks is given by
\ba
\label{eq:rNcollective}
r_{N,\text{coll}} &=& \frac{n}{N}\,\Big[\min_{E|\mathbf{V}\pm\Delta\mathbf{V}(\varepsilon_{\text{PE}})} H(A|E)\,-\,H(A|B) \,-\,\frac{1}{n}\log\frac{2}{\varepsilon_{\text{EC}}} \,-\,\frac{2}{n}\log\frac{1}{\varepsilon_{\text{PA}}}
\,-\,(2d+3)\sqrt{\frac{\log(2/\bar{\varepsilon})}{n}}
\Big]\,,
\ea 
where we are assuming that the yield of the error correction protocol is perfect, to reach the Shannon limit, $H(A|B)$.

\subsection{Beyond collective attacks}

Previous works \cite{SR08,CS09,SS10} used the bound above to claim unconditional security for the BB84 and the six-state protocols, as well as for their natural high-dimensional generalizations, because for those protocols the bound for collective attacks coincides with the one for coherent attacks \cite{KGR05}. But, for protocols using a less symmetric encoding, there is no guarantee that this is the case. The most general attacks are impossible to parametrize. Therefore, the generic recipe for unconditional security consists, in a nutshell, in bounding the possible advantage of coherent attacks over the collective ones, then computing the bound for collective attacks with the suitable overhead terms.

\begin{figure}[h!]
 \begin{center}
     \includegraphics[width=12cm]{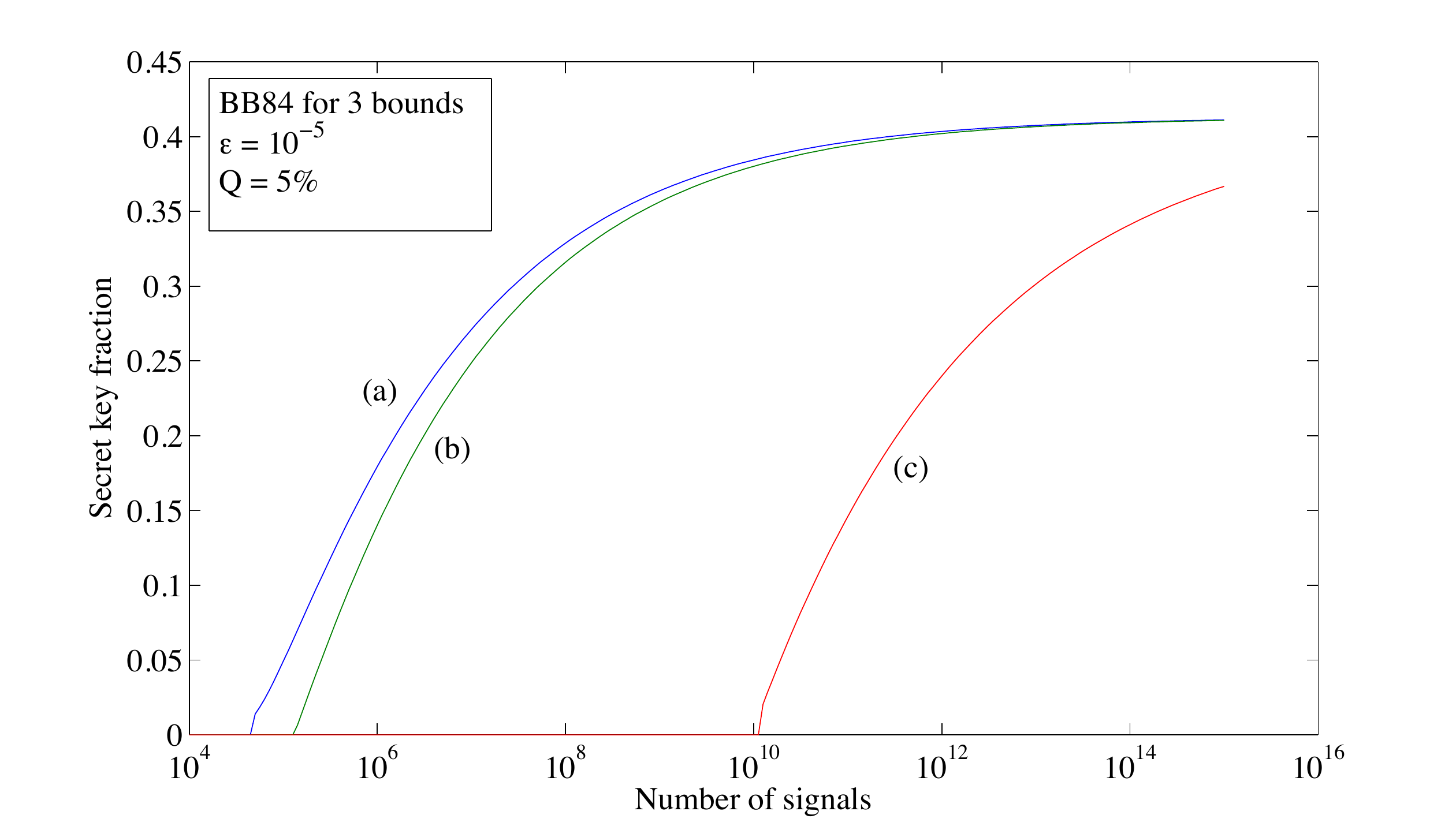}
  \vskip -1.5ex
\caption{Secret key fraction for BB84 vs. the number of signals $N$ for 3 different finite analysis bounds (a)  collective attacks, (b) the post-selection technique, and (c) the exponential de Finetti theorem.}
\label{fig:3rates}
\end{center}
\end{figure}

The first such approach used the \textit{exponential de Finetti theorem} \cite{rennerthesis,R07}. This theorem bounds the distance between any state $\rho_{AB}^{(n)}$ that leads to permutationally invariant statistics for Alice and Bob, and $n$-fold product states $\sigma_{AB}^{\otimes n}$ (or mixtures thereof), i.e. exactly the states that a collective attack would produce. The overhead obtained by this theorem turns out to be very heavy, so much so that it would make finite-key bounds unrealistically pessimistic (Figure~\ref{fig:3rates}). This fact was stressed already in \cite{SR08}, but the explicit expressions and results were not given, so we present them in Appendix~\ref{appdf}.

The de Finetti theorem is tight if one wants to compare the attacks at the level of the states. Christandl, K\"onig and Renner~\cite{CKR09} noticed that, for the sake of QKD and other quantum information processing tasks, a much less refined comparison is actually sufficient. 

\begin{figure}[h!]
\begin{center}
\includegraphics[width=6.5cm]{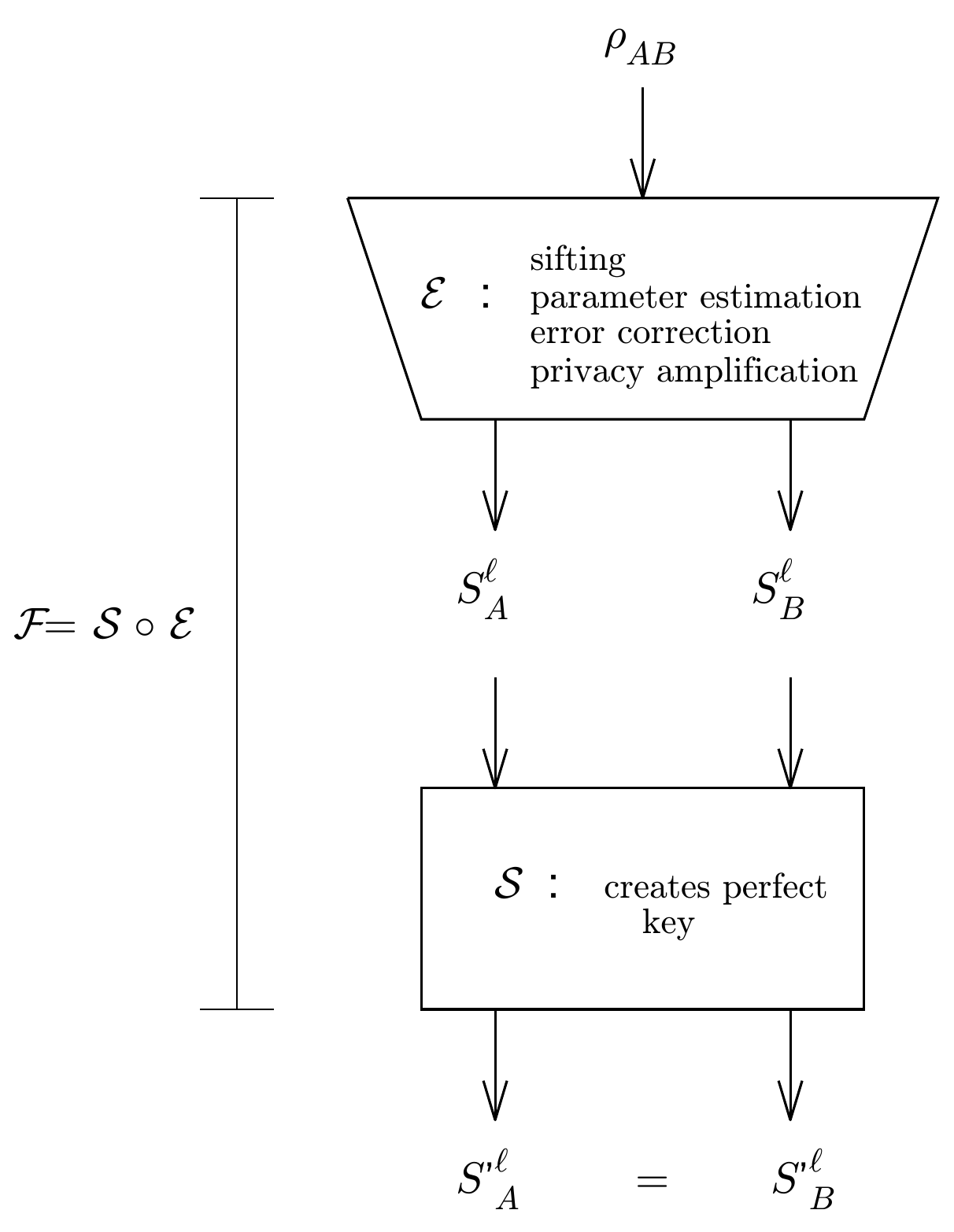} 
\caption{Consider the distance, $\Delta$, between the permutation invariant maps $\mathcal{E}$, implementing the QKD protocol, and $\mathcal{F}=\mathcal{S}\circ\mathcal{E}$, where the map $\mathcal{S}$ is a hypothetical process that takes an imperfect key to a perfect one.  This distance can be found when the maps act on the de-Finetti-Hilbert-Schmidt state, which describes the case for collective attacks, and the increase in $\Delta$ can be bounded when the same two maps act on an arbitrary state, the case for coherent attacks.  This model is from~\cite{CKR09}.}
\label{fig:CKRProtocol}
\end{center}
\end{figure}

They found that it suffices to consider the distance between two permutation invariant maps and how this distance changes when acting on states that result from a general attack rather than on states from resulting from a collective attack.   The maps are the one describing the QKD protocol being implemented and an idealized scheme which takes any quantum state as an input and distributes two classical perfectly correlated random strings to Alice and Bob.  See Figure~\ref{fig:CKRProtocol}.

In summary: let us fix $\varepsilon_{\text{coh}}$ as the tolerable failure probability of the secret key against coherent attacks. Then, the resulting expression for the secret key rate is
\ba
\label{eq:rNpostselection}
r_N &=& r_{N,\text{coll}}\,-\,\frac{2(d^4-1)\log(N+1)}{N}
\ea
where the bound for collective attacks (\ref{eq:rNcollective}) is computed under the constraint (\ref{eq:epscollective}) for the security parameter \begin{equation}
\varepsilon_{\text{col}}=\varepsilon_{\text{coh}}\,(N+1)^{-(d^4-1)}\,.\label{eq:epspostselection}
\end{equation}
The improvement that this technique gives over the use of the exponential de Finetti theorem is illustrated in Figure~\ref{fig:3rates}.  For the BB84 protocol the optimal coherent attack is a collective attack and therefore the line (a) is the best bound for security. However, if that were not known to be the case, the post-selection technique gives a bound close to the optimal one; whereas the bound obtained using the de Finetti is substantially worse and would imply the practical impossibility of obtaining a key in QKD. 

\section{Case study: reference frame independent protocol}
\label{application}

\subsection{Review of the Protocol}
\label{protocol}
We briefly describe the \emph{reference frame independent} protocol \cite{LSROB10}. In the prepare and measure scenario, Alice sends to Bob a qubit prepared in an eigenstate of three mutually unbiased bases $\{X_A,Y_A,Z_A\}$ chosen at random but not necessarily with the same probability. Bob then receives a qubit which may be tampered by Eve and measures in his own basis chosen among a possibly different set of mutually unbiased bases $\{X_B,Y_B,Z_B\}$. The equivalent entanglement based version is that Alice and Bob receive a pair of entangled qubits in a state $\rho_{AB}$ which is $|\Phi^+\rangle$ in the ideal case, and perform the local measurements defined by the above-mentioned bases on them. The measurements can be described by a vector in the Bloch sphere which we will refer to by direction. Unlike usual protocols, where the reference frames orientations are actively monitored using the classical channel, this protocol requires one well defined direction $Z_A = Z_B$ while the other two directions are related by an unknown transformation
\begin{equation}
X_B=\cos\beta X_A+\sin\beta Y_A, \ \ Y_B=\cos\beta Y_A-\sin\beta X_A\,.\label{frames}
\end{equation}
At the end of the signal exchange phase, they reveal their bases. This protocol is intrinsically asymmetric, in that the different bases play different roles. The raw key consists of the cases where both have measured in the $Z$ basis, and is characterized by the quantum bit error rate
\begin{equation} 
Q=\frac{1-\langle Z_AZ_B\rangle}{2}\,.
\end{equation}
Eve's information is quantified by the parameter
\begin{equation}
 C=\langle X_AX_B\rangle^2+\langle X_AY_B\rangle^2+\langle Y_AX_B\rangle^2+\langle Y_AY_B\rangle^2 \leq 2 
 \label{eq:Cdef}
\end{equation} where $C=2$ guarantees maximal entanglement. Note that four measurements are needed to estimate $C$, so the actual parameters that are measured are \ba
v_1=\langle X_AX_B\rangle\,,\;v_2=\langle X_AY_B\rangle\,,\;v_3=\langle Y_AX_B\rangle\,,\;v_4=\langle Y_AY_B\rangle\,. \label{eq:vs}\ea The expression (\ref{eq:Cdef}) has been chosen because it is independent of $\beta$: it retains its value even if Alice's and Bob's frames are misaligned. In the asymptotic limit, the information that Eve can gain from coherent attacks is upper bounded by \begin{equation}
\label{asymtoticeveinfo}
I_E(Q,C) = (1-Q) h\left(\frac{1+u_{max}}{2}\right) + Q h\left(\frac{1+v(u_{max})}{2}\right) 
\end{equation} where
\ba
u_{max} = \text{min}\left[\frac{\sqrt{C/2}}{1-Q},1\right] &,& v(u_{max}) = \frac{1}{Q}\sqrt{C/2-(1-Q)^2u_{max}^2}
\ea and $h(x)$ is the binary entropy. This result holds in the range $0\leq Q\lesssim 15.9\%$, which is perfectly reasonable for the quality of optical lines.

Obviously, this protocol becomes of interest if $\beta$ \textit{varies in time}: if the frames are possibly misaligned but are guaranteed to be fixed in time, one would just align them once and for all. However, it takes time to collect enough data to estimate the four average values that enter the expression of $C$: the misalignment of the frames during this time leads to a smearing of the correlations and the consequent decrease of $C$. In particular, if one waits to accumulate a very large number of signals, $C$ will ultimately drop so much that no security can be inferred: in other words, the asymptotic rate (\ref{asymtoticeveinfo}) somehow assumes not only that infinitely many signals can be collected, but also that $\beta$ is fixed. In all meaningful situations, not only the realistic secret key rate, but also the \textit{optimal} one must be determined by finite-key analysis. This is the object of what follows.

\subsection{Computing the finite-key bound}

Let us particularize the parameters that enter the finite-key bound (\ref{eq:rNpostselection}) to the protocol under study. We denote by $p_Z$ the probability that Alice and Bob choose the key basis $Z$; we assume that the other two bases are chosen with equal probability $p_X=p_Y=1-2p_Z\equiv p$. So the raw key consists of $n=N\,p_Z^{\,2}$ signals, while each of the correlators $v_j$ is estimated using $m=Np^{\,2}$ signals.

The quantity $\min_{E|\mathbf{V}\pm\Delta\mathbf{V}(\varepsilon_{\text{PE}})} H(A|E)$ is given by $1-I_E(Q',C')$ where $Q'$ and $C'$ would be the perfect estimates, which are related to the observed values $(Q,C)$ by assuming the worst case fluctuations, i.e. by increasing the error $Q$ and reducing the correlations $v_j$. Specifically, $Q'=Q+\delta(n)$ and $v_j'=v_j-\delta(m)$ where
\ba
\delta(k)&=&\sqrt{\frac{\ln(1/\varepsilon_{\text{PE}})+2\ln(k+1)}{2k}}\,.
\ea
As in previous works we us the the Law of Large Numbers as presented in Cover and Thomas, Theorem 11.2.1~\cite{coverthomas}.  Other estimates have been studied~\cite{SMU10}.  Finally, $H(A|B)=h(Q)$ where the expression is a function of the observed $Q$ and not $Q'$: the EC code must correct only the errors that have actually happened.

At present, we have everything: one just has to choose the desired security level $\varepsilon_{\text{coh}}$, give the values of $N$, $\varepsilon_{\text{EC}}$, $Q$ and $C$, then maximize $r_N$ over the other parameters under the constraints (\ref{eq:epscollective}) and (\ref{eq:epspostselection}). As anticipated, we are going to study the effects of the time variations of $\beta$.

\subsection{Dynamics of $C$ for varying $\beta$}
\label{results}

The \textit{real} evolution of $\beta$ during the protocol is, by definition, unknown: its monitoring would provide the information needed to align the frames. But in order to design a protocol and choose the suitable parameters, one must make a guess of how this evolution will be. This prior guessing is not proper to this protocol: it is a general necessity when one wants to make estimates before running the experiment (for a full discussion, see paragraph 2.3 in \cite{CS09}).

Let us start by rewriting (\ref{frames}) and (\ref{eq:vs}) as
\ba
v_1(t)\,=\,v_1(0)\cos\beta(t)\,+\,v_2(0)\sin\beta(t)&,& v_2(t)\,=\,v_2(0)\cos\beta(t)\,-\,v_1(0)\sin\beta(t)\,,\\
v_3(t)\,=\,v_3(0)\cos\beta(t)\,+\,v_4(0)\sin\beta(t)&,& v_4(t)\,=\,v_4(0)\cos\beta(t)\,-\,v_3(0)\sin\beta(t)\,.
\ea
These are the ``instantaneous values", i.e. the correlations that one would observe by freezing the frames at time $t$. Now, for simplicity we assume that the $N$ signals one is going to collect are equally spaced in time with an interval $\tau$. Then the observed correlations over the time $T_N$ required to collect the $N$ signals will be given by
$\bar{v_j}(T_N)=\frac{1}{N}\,\sum_{k=0}^{N-1}v_j(k\tau)$. In other words, denoting
\ba
\frac{1}{N}\,\sum_{k=0}^{N-1} e^{i\beta(k\tau)}&\equiv& \bar{c}_N+i\bar{s}_N\,,
\ea
the $\bar{v_j}(T_N)$ are just the $v_j(t)$ with $\cos\beta(t)$ replaced by $\bar{c}_N$ and $\sin\beta(t)$ replaced by $\bar{s}_N$. It is also easy to verify that the observed value of $C$ will be
\ba
C(T_N)&=&C(0)\,\left({\bar{c}_N}^2+{\bar{s}_N}^2\right)\,:
\ea
the quality of the initial correlations is captured by $C(0)$ and is factored out from the smearing due to the variations of $\beta$.

Let us particularize now for two possible dynamics:
\begin{itemize}

\item The frames drift apart at a \textit{constant angular velocity} $\theta(t)=\omega t$. Then
\ba
\bar{c}_N+i\bar{s}_N\,=\,\frac{1}{N}\,\frac{1-e^{i\theta N}}{1-e^{i\theta}} &\mathrm{ with }& \theta\,=\,\omega \tau\,.
\ea This leads in particular to
\ba
C(T_N)&=&C(0)\,\frac{1-\cos(\theta N)}{N^2 (1-\cos\theta)}\,.
\ea
As $\theta \rightarrow 0$, the continuous sampling limit is recovered of $C(T_N) = C(0)\frac{2 (1-\cos(\theta N))}{(N\theta)^2}$. 

\item The relative angle is following a random walk behavior: $\beta$ changes by $\pm\theta$ randomly in the time $\tau$.  One is led to compute the average value of the sine and cosine of a random walk, i.e.
\ba
\bar{c}_N+i\bar{s}_N&=& \sum_{k=-N/2}^{N/2} e^{i \theta (2k)} P_N(2 k)   \,  = \, (\cos\theta)^N
\ea
where $P_N(d)=\frac{1}{2^N}\binom{N}{(N+d)/2}$ is the probability of travelling a distance $d\in\{-N,...N\}$ in N steps.  This leads to
\ba
C(T_N)&=&C(0)\,(\cos\theta)^{2N}\,.
\ea

\end{itemize}

\begin{figure}[h!]
\begin{center}
\centerline{\includegraphics[width=9cm]{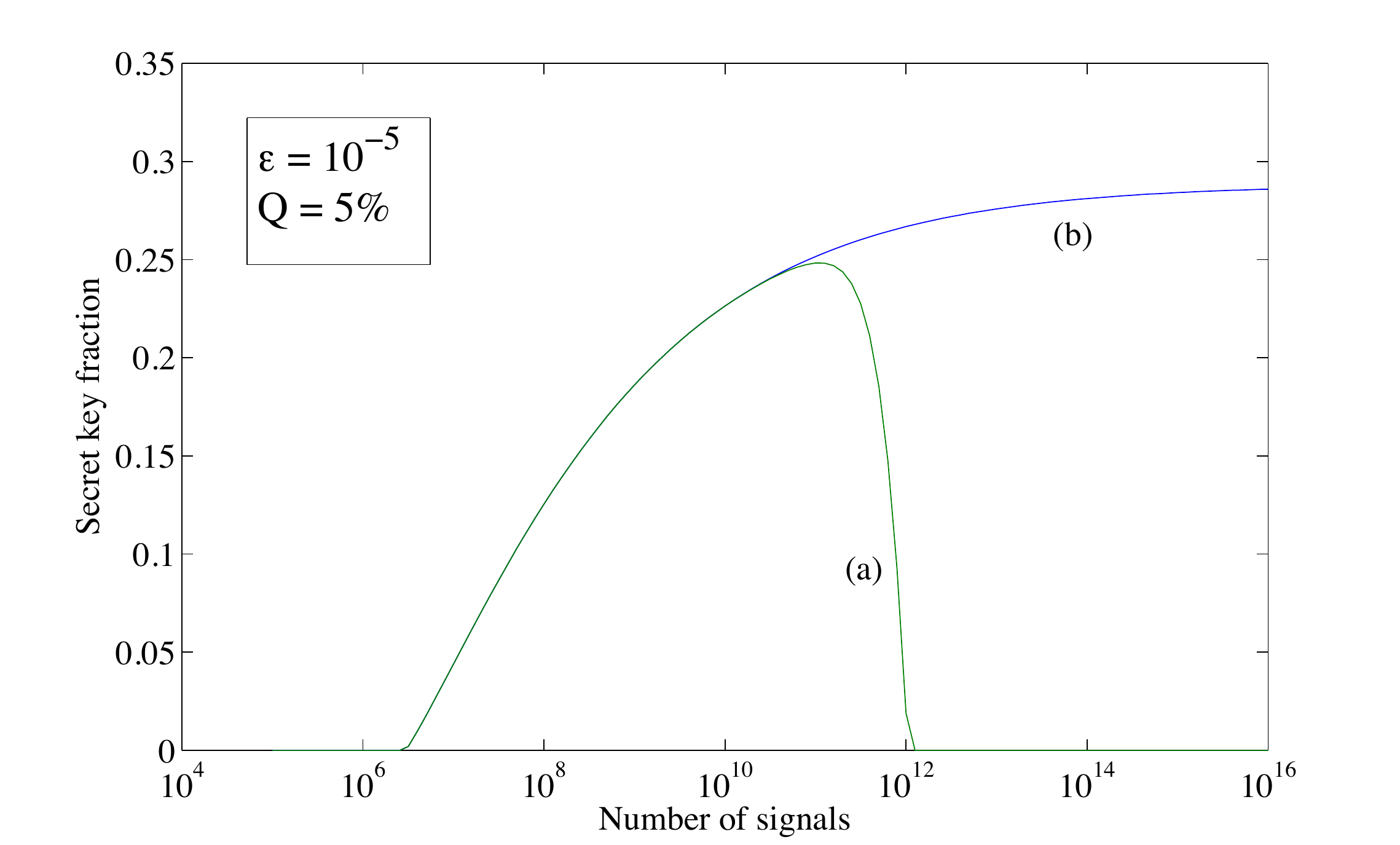} \includegraphics[width=9cm]{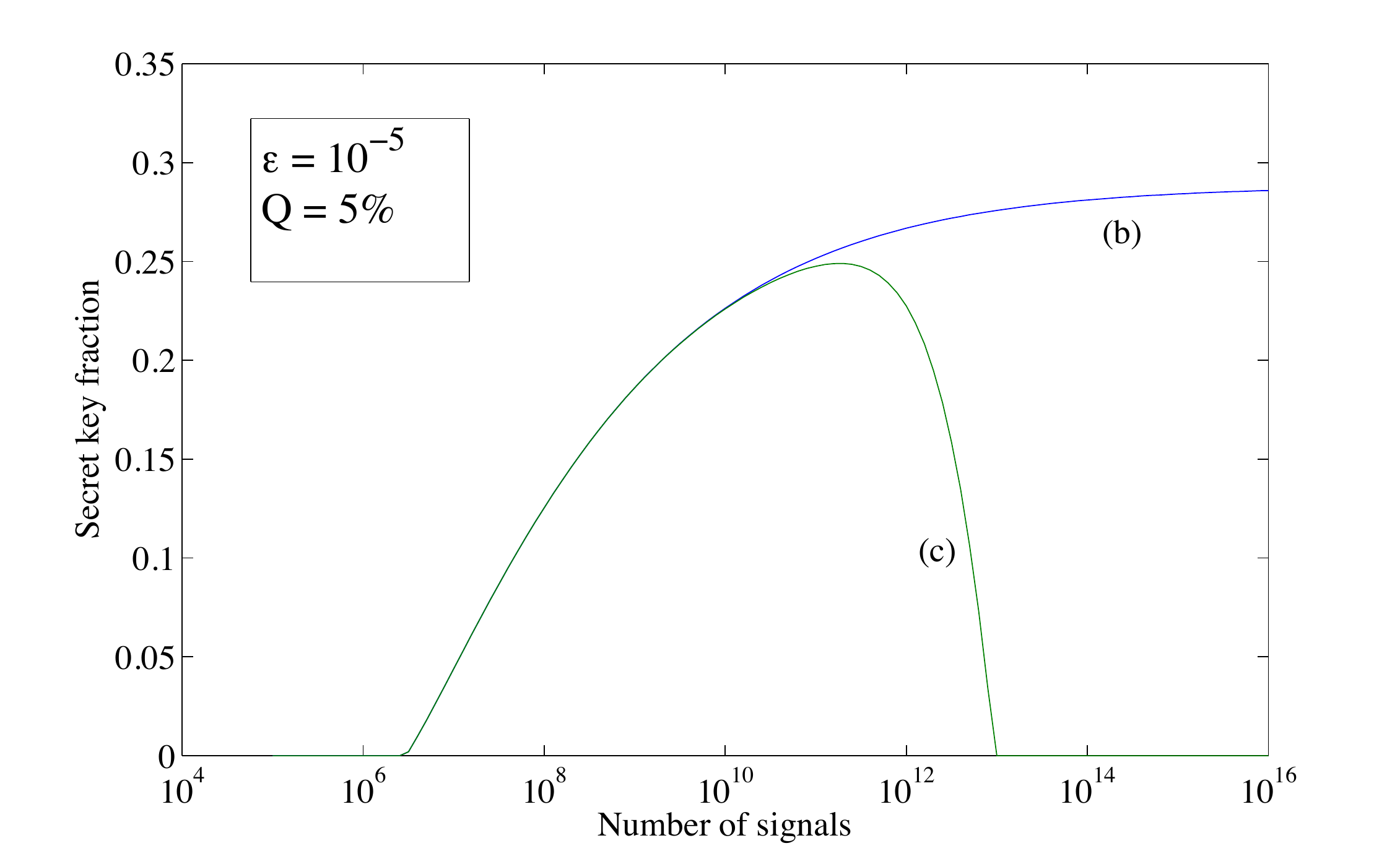}} 
\caption{Secret key fraction for (a) the frames drifting apart at constant angular velocity with $\frac{\text{d} \theta}{\text{d} N} = \frac{\pi}{180} \times 10^{-10}$, (b) fixed frames, (c) one frame drifting relative to the other according to a random walk with the different rate of $\frac{\text{d} \theta}{\text{d} N} = \frac{\pi}{180} \times 10^{-5}$ per step.  The for both plots security parameter is $\epsilon = 10^{-5}$, $C(0) = 1.72$, and $Q=5\%$.}
\label{fig:drift}
\end{center}
\end{figure}

In both cases, of course, $C(T_N)$ goes to zero for large $N$. The effect of this smearing on the finite-key secret key rate is shown in Figure \ref{fig:drift}.

\section{Conclusion}
\label{conclusion}

We have studied the application of the post-selection technique of~\cite{CKR09} to QKD protocols in finite-key scenarios to extend security bounds for collective attacks to bounds for coherent attacks.  We have compared it explicitly to the bounds recovered for finite keys using the de Finetti theorem.  We demonstrate how to compute this new bound by applying it to the \emph{reference frame independent} protocol of~\cite{LSROB10}.  In addition, we have considered two physically plausible scenarios for the case of unaligned reference frames: that one frame may be rotating relative to the other, or that one frame may be executing a random-walk-type drift relative to the other.

The most prominent feature in these two cases is that the asymptotic limit does not give the best key fraction. This can be seen in Figure~\ref{fig:drift}.  The reason is that the longer we collect the signals, the lower the value of the security parameter $C$ becomes. For a fixed $\omega$ or $\theta$, there exists an optimal block of size $N$ to obtain the best secret key fraction.  If more key is required, the protocol should be terminated and restarted after each block.  Hence any practical application of the reference frame independent protocol should aim for this optimal number of signals to be exchanged in a run of key distribution.  

\section*{Acknowledgements}

The authors would like to thank Matthias Christandl and Renato Renner for helpful discussions. This work was supported by the National Research Foundation and the Ministry of Education, Singapore.

\bibliography{QKDBib6}

\begin{thebibliography}{23}
\expandafter\ifx\csname natexlab\endcsname\relax\def\natexlab#1{#1}\fi
\expandafter\ifx\csname bibnamefont\endcsname\relax
  \def\bibnamefont#1{#1}\fi
\expandafter\ifx\csname bibfnamefont\endcsname\relax
  \def\bibfnamefont#1{#1}\fi
\expandafter\ifx\csname citenamefont\endcsname\relax
  \def\citenamefont#1{#1}\fi
\expandafter\ifx\csname url\endcsname\relax
  \def\url#1{\texttt{#1}}\fi
\expandafter\ifx\csname urlprefix\endcsname\relax\def\urlprefix{URL }\fi
\providecommand{\bibinfo}[2]{#2}
\providecommand{\eprint}[2][]{\url{#2}}

\bibitem[{\citenamefont{Gisin et~al.}(2002)\citenamefont{Gisin, Ribordy,
  Tittel, and Zbinden}}]{review1}
\bibinfo{author}{\bibfnamefont{N.}~\bibnamefont{Gisin}},
  \bibinfo{author}{\bibfnamefont{G.}~\bibnamefont{Ribordy}},
  \bibinfo{author}{\bibfnamefont{W.}~\bibnamefont{Tittel}}, \bibnamefont{and}
  \bibinfo{author}{\bibfnamefont{H.}~\bibnamefont{Zbinden}},
  \bibinfo{journal}{Rev. Mod. Phys.} \textbf{\bibinfo{volume}{74}},
  \bibinfo{pages}{145} (\bibinfo{year}{2002}).

\bibitem[{\citenamefont{Scarani et~al.}(2009)\citenamefont{Scarani,
  Bechmann-Pasquinucci, Cerf, Dusek, Lutkenhaus, and Peev}}]{review2}
\bibinfo{author}{\bibfnamefont{V.}~\bibnamefont{Scarani}},
  \bibinfo{author}{\bibfnamefont{H.}~\bibnamefont{Bechmann-Pasquinucci}},
  \bibinfo{author}{\bibfnamefont{N.~J.} \bibnamefont{Cerf}},
  \bibinfo{author}{\bibfnamefont{M.}~\bibnamefont{Dusek}},
  \bibinfo{author}{\bibfnamefont{N.}~\bibnamefont{Lutkenhaus}},
  \bibnamefont{and} \bibinfo{author}{\bibfnamefont{M.}~\bibnamefont{Peev}},
  \bibinfo{journal}{Rev. Mod. Phys.} \textbf{\bibinfo{volume}{81}},
  \bibinfo{eid}{1301} (\bibinfo{year}{2009}).

\bibitem[{\citenamefont{Bennett and Brassard}(1984)}]{bb84}
\bibinfo{author}{\bibfnamefont{C.~H.} \bibnamefont{Bennett}} \bibnamefont{and}
  \bibinfo{author}{\bibfnamefont{G.}~\bibnamefont{Brassard}}, in
  \emph{\bibinfo{booktitle}{Proceedings of IEEE International Conference on
  Computers, Systems and Signal Processing}} (\bibinfo{publisher}{IEEE},
  \bibinfo{address}{New York}, \bibinfo{year}{1984}), pp.
  \bibinfo{pages}{175--179}.

\bibitem[{\citenamefont{Ekert}(1991)}]{e91}
\bibinfo{author}{\bibfnamefont{A.~K.} \bibnamefont{Ekert}},
  \bibinfo{journal}{Phys.\ Rev.\ Lett.} \textbf{\bibinfo{volume}{67}},
  \bibinfo{pages}{661} (\bibinfo{year}{1991}).

\bibitem[{\citenamefont{Mayers}(1996)}]{may96}
\bibinfo{author}{\bibfnamefont{D.}~\bibnamefont{Mayers}}, in
  \emph{\bibinfo{booktitle}{Advances in Cryptology \,---\, Proceedings of
  Crypto '96}} (\bibinfo{publisher}{Springer Verlag},
  \bibinfo{address}{Berlin}, \bibinfo{year}{1996}), p. \bibinfo{pages}{343}.

\bibitem[{\citenamefont{Hayashi}(2007)}]{hay2}
\bibinfo{author}{\bibfnamefont{M.}~\bibnamefont{Hayashi}},
  \bibinfo{journal}{Phys. Rev. A} \textbf{\bibinfo{volume}{76}},
  \bibinfo{pages}{012329} (\bibinfo{year}{2007}).

\bibitem[{\citenamefont{Scarani and Renner}(2008{\natexlab{a}})}]{SR08}
\bibinfo{author}{\bibfnamefont{V.}~\bibnamefont{Scarani}} \bibnamefont{and}
  \bibinfo{author}{\bibfnamefont{R.}~\bibnamefont{Renner}},
  \bibinfo{journal}{Phys. Rev. Lett.} \textbf{\bibinfo{volume}{100}},
  \bibinfo{pages}{200501} (\bibinfo{year}{2008}{\natexlab{a}}).

\bibitem[{\citenamefont{Cai and Scarani}(2009)}]{CS09}
\bibinfo{author}{\bibfnamefont{R.~Y.} \bibnamefont{Cai}} \bibnamefont{and}
  \bibinfo{author}{\bibfnamefont{V.}~\bibnamefont{Scarani}},
  \bibinfo{journal}{New J. Phys.} \textbf{\bibinfo{volume}{11}},
  \bibinfo{pages}{045024} (\bibinfo{year}{2009}).

\bibitem[{\citenamefont{Sheridan and Scarani}(2010)}]{SS10}
\bibinfo{author}{\bibfnamefont{L.}~\bibnamefont{Sheridan}} \bibnamefont{and}
  \bibinfo{author}{\bibfnamefont{V.}~\bibnamefont{Scarani}}
  (\bibinfo{year}{2010}), \bibinfo{note}{arXiv:1003.5464}.

\bibitem[{\citenamefont{Kraus et~al.}(2005)\citenamefont{Kraus, Gisin, and
  Renner}}]{KGR05}
\bibinfo{author}{\bibfnamefont{B.}~\bibnamefont{Kraus}},
  \bibinfo{author}{\bibfnamefont{N.}~\bibnamefont{Gisin}}, \bibnamefont{and}
  \bibinfo{author}{\bibfnamefont{R.}~\bibnamefont{Renner}},
  \bibinfo{journal}{Phys. Rev. Lett.} \textbf{\bibinfo{volume}{95}},
  \bibinfo{pages}{080501} (\bibinfo{year}{2005}).

\bibitem[{\citenamefont{Christandl et~al.}(2009)\citenamefont{Christandl,
  K\"onig, and Renner}}]{CKR09}
\bibinfo{author}{\bibfnamefont{M.}~\bibnamefont{Christandl}},
  \bibinfo{author}{\bibfnamefont{R.}~\bibnamefont{K\"onig}}, \bibnamefont{and}
  \bibinfo{author}{\bibfnamefont{R.}~\bibnamefont{Renner}},
  \bibinfo{journal}{Phys. Rev. Lett.} \textbf{\bibinfo{volume}{102}},
  \bibinfo{pages}{020504} (\bibinfo{year}{2009}).

\bibitem[{\citenamefont{Leverrier et~al.}(2010)\citenamefont{Leverrier,
  Grosshans, and Grangier}}]{LGG10}
\bibinfo{author}{\bibfnamefont{A.}~\bibnamefont{Leverrier}},
  \bibinfo{author}{\bibfnamefont{F.}~\bibnamefont{Grosshans}},
  \bibnamefont{and} \bibinfo{author}{\bibfnamefont{P.}~\bibnamefont{Grangier}},
  \bibinfo{journal}{Phys. Rev. A} \textbf{\bibinfo{volume}{81}},
  \bibinfo{pages}{062343} (\bibinfo{year}{2010}).

\bibitem[{\citenamefont{Bennett}(1992)}]{b92}
\bibinfo{author}{\bibfnamefont{C.~H.} \bibnamefont{Bennett}},
  \bibinfo{journal}{Phys. Rev. Lett.} \textbf{\bibinfo{volume}{68}},
  \bibinfo{pages}{3121} (\bibinfo{year}{1992}).

\bibitem[{\citenamefont{Scarani et~al.}(2004)\citenamefont{Scarani, Ac\'{\i}n,
  Ribordy, and Gisin}}]{sarg04}
\bibinfo{author}{\bibfnamefont{V.}~\bibnamefont{Scarani}},
  \bibinfo{author}{\bibfnamefont{A.}~\bibnamefont{Ac\'{\i}n}},
  \bibinfo{author}{\bibfnamefont{G.}~\bibnamefont{Ribordy}}, \bibnamefont{and}
  \bibinfo{author}{\bibfnamefont{N.}~\bibnamefont{Gisin}},
  \bibinfo{journal}{Phys. Rev. Lett.} \textbf{\bibinfo{volume}{92}},
  \bibinfo{pages}{057901} (\bibinfo{year}{2004}).

\bibitem[{\citenamefont{Branciard et~al.}(2005)\citenamefont{Branciard, Gisin,
  Kraus, and Scarani}}]{BGKS05}
\bibinfo{author}{\bibfnamefont{C.}~\bibnamefont{Branciard}},
  \bibinfo{author}{\bibfnamefont{N.}~\bibnamefont{Gisin}},
  \bibinfo{author}{\bibfnamefont{B.}~\bibnamefont{Kraus}}, \bibnamefont{and}
  \bibinfo{author}{\bibfnamefont{V.}~\bibnamefont{Scarani}},
  \bibinfo{journal}{Phys. Rev. A} \textbf{\bibinfo{volume}{72}},
  \bibinfo{pages}{032301} (\bibinfo{year}{2005}).

\bibitem[{\citenamefont{Ac\'{\i}n et~al.}(2006)\citenamefont{Ac\'{\i}n, Massar,
  and Pironio}}]{AMP06}
\bibinfo{author}{\bibfnamefont{A.}~\bibnamefont{Ac\'{\i}n}},
  \bibinfo{author}{\bibfnamefont{S.}~\bibnamefont{Massar}}, \bibnamefont{and}
  \bibinfo{author}{\bibfnamefont{S.}~\bibnamefont{Pironio}},
  \bibinfo{journal}{New J. Phys.} \textbf{\bibinfo{volume}{8}},
  \bibinfo{pages}{126} (\bibinfo{year}{2006}).

\bibitem[{\citenamefont{Scarani and Renner}(2008{\natexlab{b}})}]{SR08b}
\bibinfo{author}{\bibfnamefont{V.}~\bibnamefont{Scarani}} \bibnamefont{and}
  \bibinfo{author}{\bibfnamefont{R.}~\bibnamefont{Renner}}, in
  \emph{\bibinfo{booktitle}{Proceedings of TQC2008, Lecture Notes in Computer
  Science \textbf{5106}}} (\bibinfo{publisher}{Springer Verlag},
  \bibinfo{address}{Berlin}, \bibinfo{year}{2008}{\natexlab{b}}), pp.
  \bibinfo{pages}{83--95}.

\bibitem[{\citenamefont{Laing et~al.}(2010)\citenamefont{Laing, Scarani,
  Rarity, and O'Brien}}]{LSROB10}
\bibinfo{author}{\bibfnamefont{A.}~\bibnamefont{Laing}},
  \bibinfo{author}{\bibfnamefont{V.}~\bibnamefont{Scarani}},
  \bibinfo{author}{\bibfnamefont{J.~G.} \bibnamefont{Rarity}},
  \bibnamefont{and} \bibinfo{author}{\bibfnamefont{J.~L.}
  \bibnamefont{O'Brien}}, \bibinfo{journal}{Phys. Rev. A}
  \textbf{\bibinfo{volume}{82}}, \bibinfo{pages}{012304}
  (\bibinfo{year}{2010}).

\bibitem[{\citenamefont{Lo et~al.}(2005)\citenamefont{Lo, Chau, and
  Ardehali}}]{LCA}
\bibinfo{author}{\bibfnamefont{H.-K.} \bibnamefont{Lo}},
  \bibinfo{author}{\bibfnamefont{H.}~\bibnamefont{Chau}}, \bibnamefont{and}
  \bibinfo{author}{\bibfnamefont{M.}~\bibnamefont{Ardehali}},
  \bibinfo{journal}{J.Cryptology} \textbf{\bibinfo{volume}{18}},
  \bibinfo{pages}{133} (\bibinfo{year}{2005}).

\bibitem[{\citenamefont{Renner}(2008)}]{rennerthesis}
\bibinfo{author}{\bibfnamefont{R.}~\bibnamefont{Renner}},
  \bibinfo{journal}{Int. J. Quant. Inf.} \textbf{\bibinfo{volume}{6}},
  \bibinfo{pages}{1} (\bibinfo{year}{2008}).

\bibitem[{\citenamefont{Renner}(2007)}]{R07}
\bibinfo{author}{\bibfnamefont{R.}~\bibnamefont{Renner}},
  \bibinfo{journal}{Nature Physics} \textbf{\bibinfo{volume}{3}},
  \bibinfo{pages}{645} (\bibinfo{year}{2007}).

\bibitem[{\citenamefont{Cover and Thomas}(2006)}]{coverthomas}
\bibinfo{author}{\bibfnamefont{T.~M.} \bibnamefont{Cover}} \bibnamefont{and}
  \bibinfo{author}{\bibfnamefont{J.~A.} \bibnamefont{Thomas}},
  \emph{\bibinfo{title}{Elements of Information Theory}}
  (\bibinfo{publisher}{Wiley Interscience}, \bibinfo{year}{2006}),
  \bibinfo{edition}{2nd} ed., ISBN \bibinfo{isbn}{9780471241959}.

\bibitem[{\citenamefont{Sano et~al.}(2010)\citenamefont{Sano, Matsumoto, and
  Uyematsu}}]{SMU10}
\bibinfo{author}{\bibfnamefont{Y.}~\bibnamefont{Sano}},
  \bibinfo{author}{\bibfnamefont{R.}~\bibnamefont{Matsumoto}},
  \bibnamefont{and} \bibinfo{author}{\bibfnamefont{T.}~\bibnamefont{Uyematsu}}
  (\bibinfo{year}{2010}), \bibinfo{note}{arXiv:1003.5766}.

\end{thebibliography}

\begin{appendix}

\section{De Finetti}
\label{appdf}

Here we consider the bound which can be derived from using the de Finetti bound when using $d$-dimensional systems following the results of~\cite{rennerthesis}.  

Now, of the sifted signals $N_s$, $m$ will be used for parameter estimation and $k$ systems are traced over to make use of the de Finetti theorem by bounding the remaining systems to have been very close to a mixture of product states $\sigma^{\otimes n}$.  Thus, $n = N_s - m -k$ is the number of remaining systems that can be put towards the key, but since it is not yet secure, this is the \emph{raw key}.

Let the state $\bar{\rho}^{\ket{\theta}}_n$ be the permutationally invariant output of a quantum key distribution protocol.  Because the state $\bar{\rho}^{\ket{\theta}}_n$ is in general not exactly of product form, for any $\ket{\theta}$, it is a \emph{pure} state of the symmetric subspace of $\HH^{\otimes n}$ such that $\bar{\rho}^{\ket{\theta}}_n = \sum_{\pi} \ket{\theta}^{\otimes n-t} \otimes \ket{\phi}_t$, where the sum is over all permutations, $\pi$, for some $t$ such that $0\leq t\leq m/2$.  In some sense, $t$ can be thought of as quantifying the distance that the state $\bar{\rho}^{\ket{\theta}}_n$ is from the perfect pure $n$-fold product state. 

So we can now introduce an error, $\varepsilon_{\text{deF}}$, that parameterizes $t$:
\begin{equation}
\label{eq:epscolldefin}
\varepsilon_{\text{coh}} = \varepsilon_{\text{PA}} + \bar{\varepsilon} + n_{\text{PE}}\varepsilon_{\text{PE}} + \varepsilon_{\text{EC}} + \varepsilon_{\text{deF}}.
\end{equation} 
where $t = \frac{N_s}{k}\bigl(2 \ln(2/\varepsilon_{\text{deF}})+d^4 \ln(k)\bigr)$~\cite{rennerthesis}.

The maximum error in the parameter estimation, assuming $m$ samples, is now:
\begin{equation}
\delta(m) = \frac{1}{d-1} \sqrt{(1+\ln2)\, h\left(\frac{t}{m}\right) + \frac{\ln(1/\varepsilon_{PE})+d \ln(m/2 + 1)}{m}}
%\xi = \frac{?}{?} \sqrt{h(t/m) + \frac{\log(/2\varepsilon_{PE})+d \log(m/2 + 1)}{m}}
\end{equation} 
where $k$ is optimized over.  We see then that if $k$ is larger, $t$ can be smaller (the form of the raw key state can constrain Eve to collective attacks more closely), however, this reduces the size or the raw key, so there is a trade-off.

The term giving the privacy amplification correction is also modified~\cite{rennerthesis}, so that the final rate is given by
\ba
\label{eq:rNcohdeF}
r_{N,\text{coh},\text{deF}} &=& \frac{n}{N}\,\Big[\min_{E|\mathbf{V}\pm\Delta\mathbf{V}(\varepsilon_{\text{PE}})} H(A|E)\,-\,H(A|B) \,-\,\frac{1}{n}\log\frac{2}{\varepsilon_{\text{EC}}} \\
& & \qquad -\frac{2}{n}\log(1/ \varepsilon_{PA})- \frac{2}{n}(m+k)\log(d^2) - \left(\frac{5}{2}d+4\right)\sqrt{\frac{\log(2/\bar{\varepsilon})}{n}+h(t/n)}
\Big]\,.
\ea

These expressions can be used in equation~(\ref{eq:rNcollective}) to get a bound for coherent attacks.

\section{Derivation of Eqs. (\ref{eq:rNpostselection}) and (\ref{eq:epspostselection}) from Ref.~\cite{CKR09}}
\label{appckr}

General coherent attacks can be bounded in terms of collective attacks for general permutation invariant protocols by using the method introduced in~\cite{CKR09}.

First, it is usually easier to prove that a protocol is secure against collective attacks than coherent ones, so the problem is approached for a particular state, the \emph{de-Finetti-Hilbert-Schmidt state} $\tau_{A^N B^N}$, which represents the mixture over states that could be held by Alice and Bob after Eve makes a collective attack.  This state is defined as:
\begin{equation}
\tau_{A^N B^N} = \int \sigma_{A B}^{\otimes N} \, \text{d}_{\text{HS}}\sigma_{A B}
\end{equation}
where $\text{d}_{\text{HS}}$ is the measure induced by the Hilbert-Schmidt metric, $\Delta_{\text{HS}}(X-Y) = \|X-Y\|_{\text{HS}}$ and $\|X\|_{\text{HS}}^2=\Tr(X^\dag X)$.  

Let $\mathcal{E}$ be the actual protocol for which security is to be proven and $\mathcal{F}$ be an ideal key-generation protocol composed of the actual protocol $\mathcal{E}$ and a map $\mathcal{S}$ that takes classical inputs and outputs a perfectly random perfectly correlated key string, \emph{i.e.} $\mathcal{F}=\mathcal{S} \circ \mathcal{E}$ that for any inputs gives Alice and Bob the output of an ideal key.  (See Figure~\ref{fig:CKRProtocol}.)  The main theorem of~\cite{CKR09} guarantees the security of this protocol against any coherent attack
\begin{equation}
\Delta(\mathcal{E},\mathcal{F})_{\rho} \leq (N+1)^{d^4-1} \Delta(\mathcal{E},\mathcal{F})_{\tau},
\end{equation}
where $\Delta(\mathcal{E},\mathcal{F})_{\rho}$ and $\Delta(\mathcal{E},\mathcal{F})_{\tau}$ are the diamond-norm distances between the protocols for arbitrary states $\rho$ and the de-Finetti-Hilbert-Schmidt state $\tau$ respectively, and $N$ is the number of signals or subsystems each with dimension $d^2$ (bipartite qudits shared by Alice and Bob).  Since $\rho$ is an arbitrary state it can correspond to an arbitrary quantum-mechanically-allowed attack by Eve.

\begin{figure}[h!]
\begin{center}
\includegraphics[width=11cm]{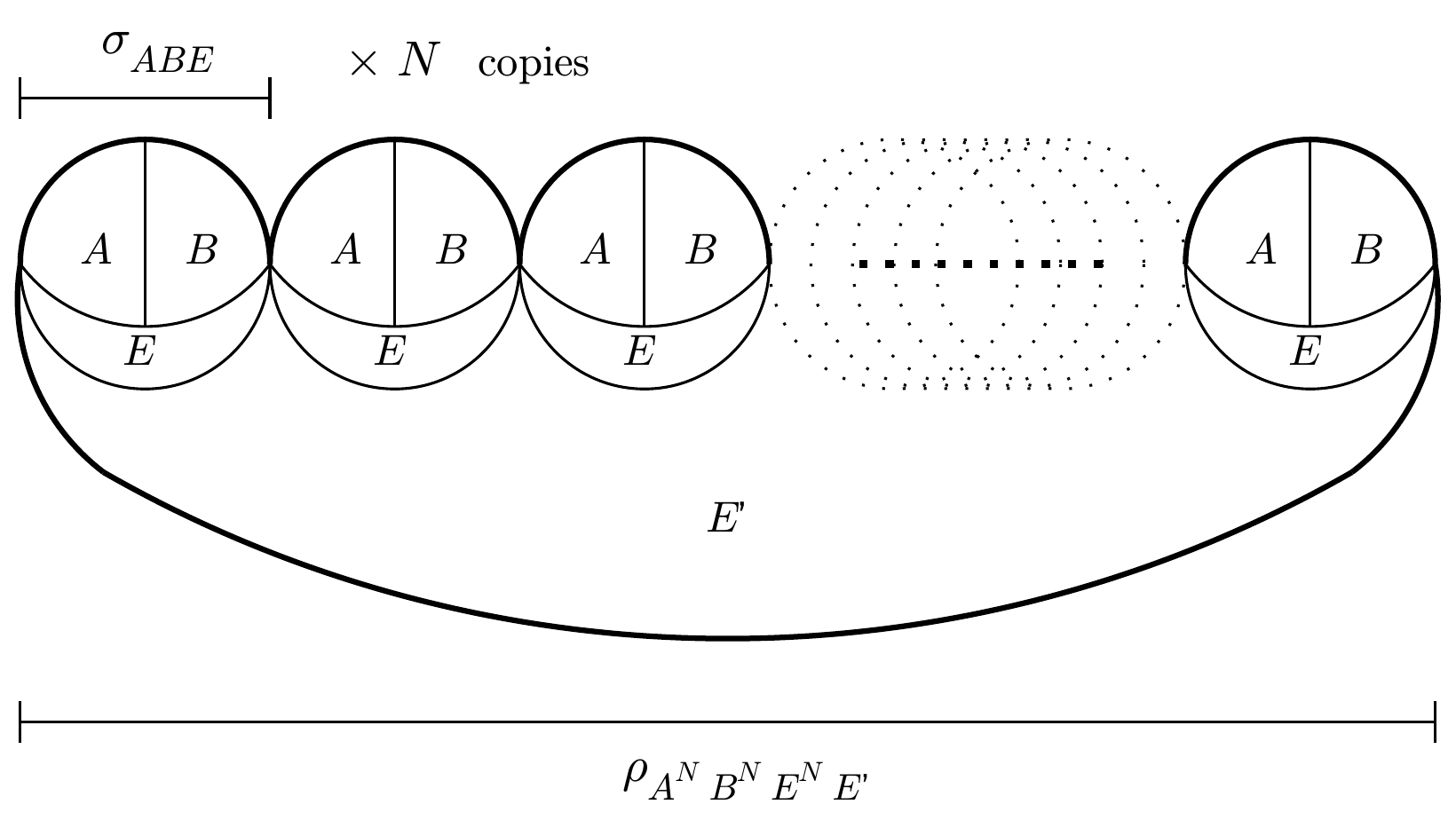} 
\caption{Eve's Hilbert space $E$ purifies each entangled system space held by Alice and Bob.  The remainder of Eve's space $E'$ purifies the state $\tau'$ on $N$ systems which is a mixture over the possible pure product states $\sigma_{ABE}$.}
\label{fig:ABEHilbert}
\end{center}
\end{figure}

In order to find the secret key fraction for finite length keys, it is also necessary to consider the effect of Eve's possession of the purification of $\rho_{A^N B^N}$.  This is already considered for collective attacks when the min-entropy of Alice's information given Eve's, $H_{\text{min}}(A|E)$, is used to bound the secret key fraction.  Let $\HH_{E}$ be the system Eve holds that purifies $\sigma_{A B}$.  (See Figure~\ref{fig:ABEHilbert}.)  Now it is necessary to also include the extra information she may have as a result of holding the purification of the mixture of the state on $N$ systems $\tau'_{A^N B^N E^N} = \int \sigma_{A B E}^{\otimes N} \, \text{d}(\sigma_{A B E})$ where $\text{d}(\cdot)$ is the Haar measure over pure states, $\sigma_{A B E}$.  Let the purification of this $N$-system state be on the Hilbert space $\HH_{E'}$.  So now we must consider $H_{\text{min}}^{\bar{\varepsilon}}(A^N | E^N E')$ in the equation for the secret key fraction.  We use the entropy bound
\begin{equation}
H_{\text{min}}^{\bar{\varepsilon}}(A^N | E^N E') \geq H_{\text{min}}^{\bar{\varepsilon}}(A^N | E^N) - 2 H_0(E') .
\end{equation}
A space of dimension no more than $(N+1)^{d^4-1}$ is needed to construct such a purification and so $\HH_{E'}$ cannot contain more than $\log \big[(N+1)^{d^4-1}\big]$ bits of information.  We therefore subtract twice this from the available entropy and divide by the number of signals $N$ to obtain equation~(\ref{eq:rNpostselection}).

So, the post-selection technique gives another way to relate a bound that can be shown for collective attacks to a bound for an unknown optimal coherent attack, provided that there is a bound on the dimension of the systems being exchanged $d$.  In other words, this result just as the de Finetti theorem cannot be used as such for continuous variables.

\end{appendix}

\end{document}